# Some things to know about achieving artificial general intelligence

Herbert L. Roitblat

Abstract

Current and foreseeable GenAI models are not capable of achieving artificial general intelligence because they are burdened with anthropogenic debt.  They depend heavily on human input to provide well-structured problems, architecture, and training data.  They cast every problem as a language pattern learning problem and are thus not capable of the kind of autonomy needed to achieve artificial general intelligence.  Current models succeed at their tasks because people solve most of the problems to which these models are directed, leaving only simple computations for the model to perform, such as gradient descent.  Another barrier is the need to recognize that there are multiple kinds of problems, some of which cannot be solved by available computational methods (for example, "insight problems").  Current methods for evaluating models (benchmarks and tests) are not adequate to identify the generality of the solutions, because it is impossible to infer the means by which a problem was solved from the fact of its solution. A test could be passed, for example, by a test-specific or a test-general method.  It is a logical fallacy (*affirming the consequent*) to infer a method of solution from the observation of success.

Predictions that artificial intelligence is on the verge of achieving general intelligence keep on coming (e.g., Altman, 2025; Leike & Sutskever, 2023; Modei, 2024).  For example, one estimate is that 88% of the necessary capabilities (Thompson, 2025) have been achieved.  The AI safety clock (International Institute for Management Development, 2024) is a symbolic representation to how close the world is to not just artificial general intelligence, but uncontrolled AGI. OpenAI recently announced a "breakthrough" on the  ARC-AGI benchmark (Chollet, 2019).

These predictions have also raised widespread concerns.  The Future of Life Institute (2023) currently has over 33,000 signatures on its open letter to temporarily halt the development of large language models. The Center for AI Safety (2023) released a statement, signed by hundreds of individuals, that "Mitigating the risk of extinction from AI should be a global priority alongside other societal-scale risks such as pandemics and nuclear war." The Artificial Intelligence Action Summit convened in Paris in February 2025.

On the other hand, Lu et al. (2024) argue that large language models only follow instructions and have no potential to autonomously master new skills without some explicit instruction.

Governments around the world are considering regulations based on these threat assessments.  For example, in 2024, the California legislature passed and the governor vetoed SB-1047 Safe and Secure Innovation for Frontier Artificial Intelligence Models Act. (2023-2024).  Among other things, it required "implementing the capability to promptly enact a full shutdown" of any covered AI model—a "kill switch".

According to a study commissioned by the US State Department (Harris, Harris, & Beal, 2024) (The Gladstone Report),

*The recent explosion of progress in advanced artificial intelligence (AI) has brought great opportunities, but it is also creating entirely new categories of weapons of mass destruction-like (WMD-like) and WMD-enabling catastrophic risks [1–4]. A key driver of these risks is an acute competitive dynamic among the frontier AI labs that are building the world's most advanced AI systems. All of these labs have openly declared an intent or expectation to achieve human-level and superhuman artificial general intelligence (AGI) — a transformative technology with profound implications for democratic governance and global security — by the end of this decade or earlier [5– 10].*

My intention is not to debate directly the potential future risks of AI, rather it is to examine the current class of models and their potential to achieve artificial general intelligence. The current models are too basic and too dependent on human intervention to pose the kind of risks that governments and others are concerned about.  General intelligence may be achievable, but it will require inventions, techniques, and discoveries that are yet to be made, and which are very difficult to predict.

Knowing how current models work, and their prospects for leading to general intelligence is critical. Regulations based on misunderstanding will be either ineffective or harmful.  Similarly investment strategies based on exaggerated claims are likely to fail.

In my opinion, the foreseeable risk from AI comes not from intelligence or superintelligence, but from the opposite, from stupidity, and from how people may misuse that stupidity in combination with people's credulity. The risk is not from what computers can do, it is from what they cannot do, and from what we mistakenly believe that they can do.

## Artificial General Intelligence

The goal of artificial general intelligence is to build a model that is capable of the full range of human problem solving.  The models do not have to solve all problems in the same way that humans do, but if they are to be considered general intelligence, they need to cover the full range of problems.

John McCarthy, who coined the term "artificial intelligence" defined it in a [proposal](proposal) for a summer workshop on the topic (McCarthy, et al., 1955).  "The study is to proceed on the basis of the conjecture that every aspect of learning or any other feature of intelligence can in principle be so precisely described that a machine can be made to simulate it."

I.J. Good (1965), taken by many to be the proposer of an eventual AI singularity, defined superintelligence this way: "Let an ultraintelligent machine be defined as a machine that can far surpass all the intellectual activities of any man however clever."

Even before that, Herbert Simon (Simon & Newell, 1958) defined general intelligence in a keynote speech:

"It is not my aim to surprise or shock you—but the simplest way I can summarize is to say that there are now in the world machines that can think, that can learn and that can create. Moreover, their ability to do these things is going to increase rapidly until – in a visible future - the range of problems they can handle will be coextensive with the range to which the human mind has been applied."

## Range of problems

By the end of the 1960s, the quest for general intelligence was predominantly replaced by a more pragmatic approach. Rather than try to solve everything all at once, the focus narrowed to solving specific problems, and then specific problems with economic value, for example, detecting cancer in radiologic images.

Although people, such as Douglas Hoftsatadter (1979) argued that such problems as winning at chess would require general intelligence, instead, the most successful chess programs were narrowly focused on traversing a tree of potential moves and counter moves.  The problem of playing chess was reduced to the problem of traversing a tree, a much simpler task for which computational methods were available.

The problems on which artificial intelligence was successful were all well-structured.  A well-structured problem (Simon, 1973) includes a definite criterion for testing a potential solution, a problem space representing the initial problem state, the final goal state, and all intermediate states, and a set of potential state changes (moves). The moves could be traversing down the branches of a tree (as in chess playing) or adjusting a parameter, for example.  Solving a problem is the process of finding a sequence (path) of moves to achieve an explicitly defined outcome.

Games like chess and go are well-structured and also perfect information problems.  There is no uncertainty about the current state of the game, it is defined perfectly by the position of the pieces on the board.  There is no randomness.  The opponent's moves depend on the player's choices not the roll of dice or other, unpredictable event.

Games like chess and go are also formal problems.  The legal moves of chess and the consequences of making them do not depend, for example, on the physical properties of the chess pieces and the board.  Playing chess does not depend on physically moving the pieces.  Entire games of chess can be played symbolically, without ever touching a physical piece.  As a result, computers can play one another without any human intervention because the rule and the states can be perfectly specified.

Narrow AI has been remarkably effective at solving a range of these well-structured problems.  But many natural problems are unstructured or ill-structured.  There is no predefined problem space, no fixed set of moves, no clearly defined and easily verified goal.  The state of the problem may not be clearly known at any point in time, there may be uncertainty, and the consequences of any particular move may be not be known for some time.

Natural problems may also be vexatious.  For example, according to the World Economic Forum (2024), Global Shapers Community Report 2022-2023, the priority problems for the world include:

- Combatting climate change and safeguarding nature
- Closing skills gaps and driving ethical innovation
- Strengthening civic spaces and diversifying leadership

- Advancing diversity, equity, inclusion, belonging and social justice
- Safeguarding mental health and overcoming persistent disparities
- Responding to disasters and in times of crisis

These problems are ill-structured and under-specified. They have no definitive formulation by which they could be solved. We may have hypotheses, but we do not know how to go about solving them. It is difficult to determine whether a particular action serves to approach a solution.

A major part of a vexatious problem, or any ill-structured problem, is just framing it in a useful way and determining what approaches to try. Is climate change best viewed as a problem of reducing carbon in the atmosphere (for example by sequestering it), or one of reducing the introduction of new carbon? Is it better to build projects that mitigate the effects of climate change, such as the [barrier installed to protect Venice](#) (Moraca, 2024) or is it better to reduce the carbon? Is it better to raise Venice or install flood barriers at the entrance to the lagoon? One key problem (meta problem?) in addressing these is identifying what exactly are the specific problems we need to solve and what are the means by which we will attempt to solve them. Even after attempting to solve them, there is no way to definitively determine if the solution has been effective.

Artificial general intelligence should be able to address problems and meta-problems like these. Perhaps it might be too much to expect any model to address issues of the magnitude described in the Global Shapers Report, but analogous, limited-scope, problems are faced every day by ordinary humans and so should be addressable.

When AI methods have been applied to ill-structured problems, the approach is usually for a human to simplify the problem into a well-structured one. Humans apply enough structure to the problem to leave just a well-structured one for the computer to solve. Once the problem is structured, then it can be solved using well-known computational techniques (e.g., gradient descent algorithms). But solving this simplified problem may not provide an actual solution to the full problem.

Even well-structured problems presently require substantial human input. The GenAI models are capable of addressing many nominal problems, but that is because all of the truly difficult parts of the problem have been provied by human designers. For example,

Human contributions to GenAI problem solving:

- Training data
- Number of neural network layers
- Types of layers
- Connection patterns
- Activation functions
- Training regimen for each layer
- Number of attention heads
- Parameter optimization method
- Context size

- Representations of words as tokens and vectors
- Training task
- Selection of problems to solve
- Training progress measures and criteria
- Human feedback for reinforcement learning
- Rules for modifying parameters as a result of human feedback
- Prompt
- Temperature and other meta-parameters

Machine contribution to GenAI problem solving:

- Parameter adjustments through gradient descent

The need for human input and the role it plays in solving these problems is rarely acknowledged or reported. For example, Goh et al. (2024) studied the diagnostic capabilities of what they called "the standalone performance of the LLM" compared with the performance of two groups of physicians, one of which had access to a large language model, and the other of which did not. Later in the paper, they mention that they spent significant time to iteratively develop a zero-shot prompt for the standalone version of the language model. So, what they were actually comparing was an LLM with a carefully designed prompt versus the same model with prompts created in a few minutes by physicians unskilled in prompt design. The difference between groups, then, was the quality of the prompt, not the quality of the model used to make the diagnosis. The differential intelligence was again provided by the prompt writer; the model was not effective without the carefully constructed prompts.

Models with this level of dependence on humans cannot be autonomous, cannot escape control, and cannot be generally intelligent. An AI model cannot achieve the goals of general intelligence until it can address the full "range [of problems] to which the human mind has been applied" (Simon & Newell, 1958), from inception to specification of the problem space to finding an appropriate solution and verifying it. The kinds of problems that humans now solve for AI must surely be within the range described by Simon and Newell.

Once the human contribution is provided, all that is left is for the models execute conceptually simple "moves" to adjust the values of the provided parameters to meet the provided goal of the computation.

Many of the problems that humans solve for the AI models can be described as "insight" problems. Multilayer neural networks became practical when someone (actually several people over time; Schmidhuber, 2022) had the insight to adjust parameters through "backpropagation," error correction proportionate to a parameter's contribution to the observed error. Before that, it was well known that single-layer perceptrons (simple neural networks) could learn to solve certain problems and the algorithm for doing that was known (the perceptron learning rule). It was also known that multi-layer perceptrons could solve more complex problems, but there was no known algorithm to use to train them.

More generally, insight problems cannot be solved by a step-by-step procedure, like an algorithm. Specific instances of a problem may be solvable by "brute force," that is iteratively trying many potential solutions, but a general solution is possible through a kind of restructuring of the solver's approach to the problem. Unlike well-structured problems, it is not possible to assess whether each step brings the solver closer to the goal. Until the right structure is achieved, all solutions are wrong.

One example of an insight problem is the [mutilated checkerboard](#) problem (Black, 1946; Heule, Kiesl, & Biere, 2019). A regular checkerboard has 32 black squares and 32 red squares. If we had 32 dominoes, each the size of two squares, it would be obvious that we could cover a checkerboard with those 32 dominoes. If we cut off the red square at the upper left corner of the checkerboard and the red square in the lower right corner of the checkerboard, could we now cover the 62-square mutilated checkerboard with 31 dominoes? There are still twice as many squares as dominos, so it may be solvable.

One potential solution would be to treat it as a well-structured problem and successively try possible arrangements of the 31 dominos. If the first arrangement does not work, the second might, and so on. A full checkerboard has over 12 million ways to arrange the dominos. So, for a human, this brute force approach would not be feasible.

Being faster, a computer might be able to try all possible arrangements, but it would be a lot faster if the computer had the insight that each domino must cover one square of each color. After the mutilation, there are now more black squares than red squares, so it is not possible to cover the mutilated checkerboard. This insight is called a "[coloring argument](#)" ([Ardilla & Stanley, undated](#)) and it is very commonly used to solve many tiling problems. So, this insight has applications far beyond the original mutilated checkerboard, but the brute force method can only solve this specific problem. A larger checkerboard would have to be solved with the same brute force method, but the coloring argument could determine immediately whether it could be covered. Once we think of it in the right way, the solution to the problem becomes obvious.

Kaplan and Simon (1990, p. 2) describe insight problems with this metaphor:

"Imagine that you are searching for a diamond in a huge, dark room. … One option is to grope blindly in the dark. … But after groping blindly for several minutes, you might decide to abandon the search for the diamond, and to search instead for a light switch. If one could be found, and the light turned on, the location of the diamond could be evident almost at once."

Current AI models can grope along the floor, or they can search for a light switch, but unless both strategies are designed into them, they cannot switch from the approach they were designed for to another one.

Current GENAI models can report an insight solution to known problems, because descriptions of the solutions to problems are widely available on the Web. They can report solutions that people have found and written about. They would not, however be able to solve insight problems for which there is no known answer, but that is a skill that a general intelligence would need.

General intelligence is itself an example of an ill-structured problem. It cannot be described with an explicit set of moves, because a necessary characteristic of general intelligence is that it create that set of potential moves. It cannot select from an explicit set of moves because that set has not been specified. The value of each move cannot be assessed quantitatively, because we do not have a quantitative way to estimate how close we are to achieving general intelligence. Human-like intelligence tests are not valid for assessing machine intelligence for reasons to be taken up in the next section.

## Measures of general intelligence

Although there are many postings and articles that claim to measure progress in artificial general intelligence (e.g., Thompson, 2025; Wade, 2024, Dilmegani, 2024), there is little justification for their estimates. Because general intelligence is an ill-structured problem and because it is general, measuring it cannot be reduced to simple tests and benchmarks. Tests and benchmarks demonstrate that a model has achieved a certain level of behavior, but they cannot be used to infer how that level was achieved.

### Human intelligence tests are unsuitable to machine intelligence

When intelligence tests are administered to people, the assumption is that the test taker has human intelligence. Colloquially, the question is how much of that human intelligence does a specific individual have.

The Paris school system asked Alfred Binet at the start of the 20th Century to find a way to [identify students who would need](#) help to receive an effective education (van Hoogdalem, & Bosman, 2024). Binet and Theodore Simon looked for indicators of the constituents of intelligence, such as language skills, memory, reasoning, the ability to follow directions, and the ability to learn. They could not test an individual's intelligence directly, rather they needed proxies or surrogates of intelligence that they hoped would be independent of the amount of schooling the child had received.

Many other intelligence tests followed, but they all had the same property of evaluating behaviors that are correlated with intelligence and are easy to score. For example, [vocabulary size](#) has some relation to general intelligence (Terman, et al., 1918). Either smart people use more unique words or using more unique words contributes to their score on the tests, or it could be correlated for other reasons. So, one of the measures of intelligence, could be the child's vocabulary size as estimated by a vocabulary test.

Intelligence tests, when given to humans, suffer from substantial biases, because the proxies that are measured are inherently culture-related. For example, it would be absurd to estimate a child's intelligence using an English vocabulary test if the child had been raised in a Spanish speaking environment (Terman, et al., 1918). Asking a child to identify a telephone in a picture would require a very different picture today than it would have in the 1950s. Other factors also risk biasing the results, but scores on intelligence tests are moderately correlated with one another and with some other measures of success.

These tests are wholly inappropriate when used to assess machine intelligence. Rather than asking how much intelligence a machine has, we need to ask whether the machine has intelligence. The vocabulary of the model is determined completely by the designer, so it would be absurd to use vocabulary as an estimator of intelligence. The "experience" of the model is similarly determined by the designer, so most other proxies for human intelligence would not be useful. Humans acquire their knowledge as a result of their experiences, but the computer is trained on predetermined collections of texts that are orders of magnitude larger than a human's. Human memory is fallible, but the computer has perfect memory (though often not verbatim). According to the College Board, 20 hours of practice on the SAT (Scholastic Aptitude Test) produces a 115-point improvement in SAT scores. Current LLMs have access to far more practice material and previous tests than a human could memorize during those 20 hours. The use of aptitude tests, then is equally absurd.

Intelligence tests have limited validity (Gygi et al., 2017) for predicting human success, such as grades. That is, they are significant, but relatively weak predictors of what a child's grade average will be in the years following a test. They have no validity for machines because the training set includes many similar tests and the source material used to design the tests, the memorization of which, would be sufficient to score highly on the test. And the training process eliminates the correlations between the characteristics that would be predictive in humans and the measure of success.

There is no *a priori* reason to think that computers have human-like intelligence and so it would require strong evidence to conclude that they do. That evidence cannot be obtained from human intelligence tests. Because the basic assumptions of those test are not met by machine systems.

## Benchmarks are unsuitable measures of general intelligence

It is customary, or at least common, to evaluate AI models against one or more benchmarks. A benchmark is a collection of problems along with their correct responses. Benchmarks allow different models to be compared on the same set of tasks and improvements on the benchmark can be seen as approaching the ultimate solution to the class of task.

One of the first benchmarks for artificial general intelligence was, of course, the Turing test (Turing, 1950). A conversation was seen as distinctive evidence of intellectual prowess and because it could be about anything, it was also seen as evidence of generality. Progress on this test could be scored, for example, as the number of minutes of conversation until the judge decided that she or he was conversing with a computer.

But, as Ned Block (1981) pointed out, the ability to hold a conversation is not a guarantee of intelligence. Other means could be used to generate the conversational language and depending on people's willingness to believe is a weak method at best. I will return to this theme shortly.

More currently, another of the benchmarks for general intelligence is the Winogrande challenge (Sakaguchi, Le Bras, Bhagavatula, and Choi, 2019). It is based on an earlier benchmark called the Winograd Schema Challenge (WSC) (Levesque, Davis, and Morgenstern 2011). These benchmarks are intended to be tests of commonsense reasoning when interpreting sentences with ambiguous pronouns.

For example, who does the word "they" refer to in the two following sentences?

- The city councilmen refused the demonstrators a permit because they feared violence.

- The city councilmen refused the demonstrators a permit because they advocated violence.

In the first sentence, "they" refers to the councilmen. In the second, "they" refers to the demonstrators. The intention of this benchmark was to be a task that required commonsense and could not be solved (they thought) by statistical language models. It takes commonsense to know, for example, that the council would make decisions based on what the council members feared, but would not make a decision based on anything that the demonstrators feared. In the second sentence, common sense would indicate that the council would be concerned about what the demonstrators advocated, but the council members would not be concerned with what the council advocated. The structure of the two sentences is otherwise the same.

One problem with the Winogrande benchmark is that there is no guarantee that the problem really does require commonsense, that it cannot be solved by statistical biases. For example, the word "fear" might be more strongly associated with councilmen, and the word "advocated" might be more strongly associated with demonstrators.

In order to be manageable, benchmarks focus on a narrow range of capabilities. The Winogrande benchmark focuses on commonsense reasoning, but only a narrow part of commonsense. Further, commonsense reasoning is only one potential part of general intelligence. Many other capabilities would also be required, but these would be outside the scope of any benchmark.

Benchmarks cannot evaluate the generality of intelligence because of their limited scope. They cannot evaluate the ability of an agent to create new insights, because they are limited to questions with known specific answers.

Benchmarks are appropriate when the task of the benchmark is the goal, for example, to correctly read radiographic images. They fail, however, when they are used to infer more general properties or infer the cause of success. Benchmarks operationalize artificial general intelligence to a low common denominator that is a reduced subset of the general capabilities.

As Block (1981) argued, demonstrating that a system can perform a task is not sufficient to declare that the system is an artificial general intelligence. More generally, when testing for intelligence, the question is not "can a system produce correct answers?", it is "does the system use general intelligence to produce its answers?" Benchmarks are designed to answer the first question, but they fail to address the second one. Benchmarks are of no value when trying to infer the cause (general intelligence) from its effect (getting the answer correct). There are many ways by which a correct answer can be guessed, general intelligence is just one of the hypothetical methods. But, asserting that a hypothesis is true from observing the results is a logical fallacy of [affirming the consequent](). It is not a valid form of reasoning. Knowing that the cookies are missing from the cookie jar does not reveal who took the cookies. Passing a test does not say whether the test taker knew the answers or just knew the answer key.

Once a benchmark has been publicly described it is likely to be included in the training set of subsequent GenAI models. The answers, therefore, could be memorized and success on them would not demonstrate that cognitive capabilities, such as commonsense reasoning, what the benchmark was intended to measure, is the cause for the correct answer. The benchmark task could be solved as a classification problem, rather than as a result of a cognitive skill or general intelligence.

As mentioned, the validity of the benchmark could also be questioned. The Winogrande challenge assumes that probabilities could not explain the results, but that assumption has not been fully validated.

Finally, general intelligence is general. Benchmarks are specific. No set of benchmarks, each of which with specific problems and known answers, can be sufficient to demonstrate this generality. A critical property of general intelligence is the ability to innovate in unanticipated ways. It would be very surprising if there could be a benchmark for this kind of innovation.

## Success on a test or benchmark does not provide information about the cause of that success

As has already been mentioned, success on a test or benchmark does not guarantee that the problem has been solved in the intended way. More generally, one cannot infer the mechanism by which a model succeeds from the fact of its success.

There are many models that have been tested on the Winogrande challenge (Papers with Code, no date). There are ten models with accuracy above 85%. One of those happens to be the leader, but they all succeed with substantial accuracy and they use different models to achieve it. Knowing that a model scored above 85% does not tell us which method it used.

Another benchmark intended to provide a measure of general intelligence, is ARC-AGI (Chollet, 2019). This benchmark is a grid-based graphic inference test. The problem consists of two or more pairs of demonstration grids. Each pair demonstrates a rule by which the first grid in the pair is transformed into the second grid. Each grid can be up to 30 rows and 30 columns, with each cell colored with one of 10 colors. Given the demonstration examples, the correct solution is to instantiate the same rule in the output grid of a test pair (Figure 1)

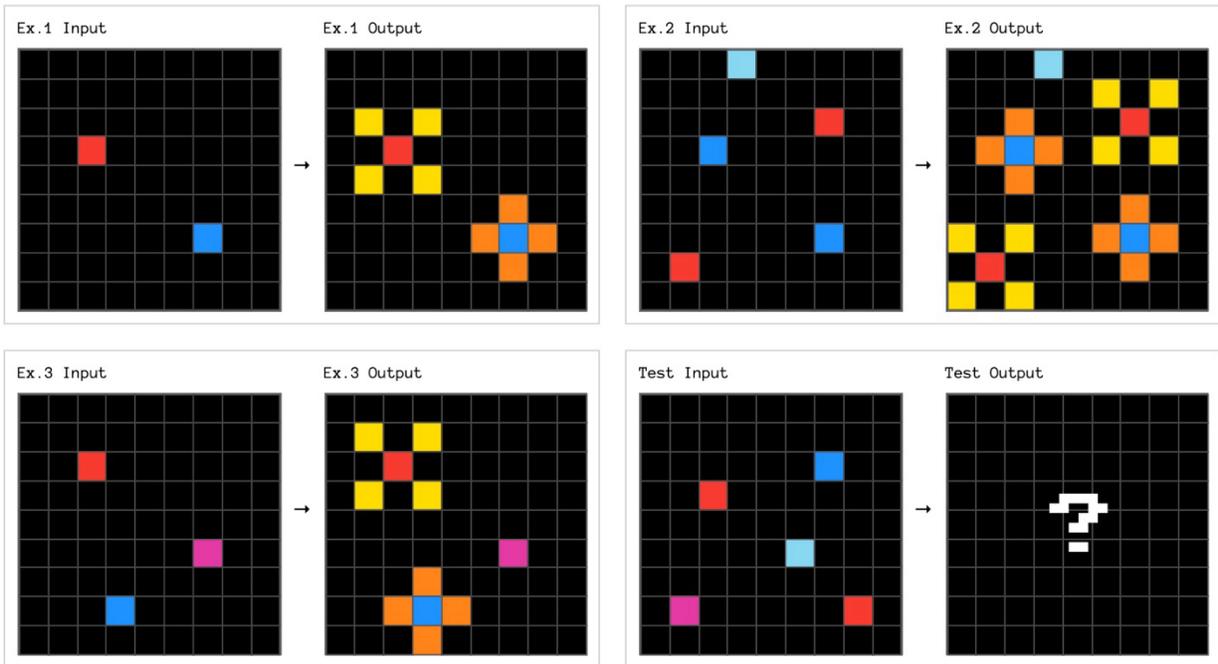

*Figure 1. Source: https://arxiv.org/html/2412.04604v1 ARC Prize 2024: Technical Report*

ARC-AGI is intended to be a test of abstraction and reasoning, which is needed for general intelligence. Assuming (for sake of argument) that it is such a test, is it the model that is doing the reasoning?

The concepts it tests are intended to be open-ended. Open ended means that there is no predefined set of concepts, but in this test, the concepts to be examined are selected from a small set of core knowledge:

- Objectness and elementary physics
- Agentness and goal-directedness
- Natural numbers and elementary arithmetic
- Elementary geometry and topology

And they are selected to be sensible to humans so the problems can be "easily" solved by them. With these constraints, there may not be many patterns that would have to be learned compared to the combinatorics of language modeling. Chollet (2019), the creator of the test, describes examples of the kinds of transformation that these concepts would allow. There may be too many potential patterns for a human to use this strategy to solve the puzzles, but models with billions of parameters should be able to handle them.

Beside open-ended concept learning, an alternative strategy to solving the ARC problems could be to generate all possible grid patterns that would fit the constraints, and then have human labelers label those that would be consistent with the core knowledge. The labelers could then further label each one with the potential transformations that could be demonstrated by that pattern, and another computer program could apply those transformations for the labelers to verify. Now each pattern is labeled for its transformation. The pattern of each example is abstracted by the labelers and the reasoning is then also supplied by the labelers. The labelers provide the abstraction and the reasoning, and the model just reports what the people have decided for each pattern. The end performance under this scenario is a combination of human intelligence and machine reporting, not evidence for model reasoning at all.

We cannot determine from success on this benchmark or any other how that success was achieved. The alternative scenario I describe is plausible and so, despite good intentions, this benchmark does not provide definitive evidence of progress toward artificial general intelligence.

## Simplifying complex ill-formed problems so a computer can solve them is not solving the complex problem

The most famous equation in the world, $E = MC^2$, can be solved by anyone who passed high school algebra, but in formulating it, Einstein revolutionized physics by showing how energy and mass could be unified ([Fernflores, 2019](#)). Einstein's insight was a revolution in physics. In Newtonian physics, inertial mass was conceived as an intrinsic property of an object. Einstein's insight unified mass and energy as equivalent, related by a constant. The problem of formulating the equation is different from and far more complex than the problem of solving it. Formulating the equation required a new way of representing mass and energy that was inconsistent with the common thinking of the time.

While problems like combatting climate change and revolutionizing physics may be too grand for artificial intelligence research, there are other, less complex, but still ill-structured problems that are suitable. One of these is the set of Bongard problems.

Bongard (1970) described an original set of 100 visual concept identification problems as part of an investigation into computer vision systems. Each problem presents two sets of six simple line drawings, each. A certain feature is shared by each of the drawings in the first set but is absent from the drawings in the second set. The task is to specify what that feature is. Foundalis (undated) catalogs [394 Bongard problems](#) and because they could employ any visual relational concepts, more could be generated. [Nie, et al. (2020)](#) claim a benchmark of 12,000 Bongard-type problems.

There are two general approaches that could be used to solve these problems. One of these would be a general, open-ended visual categorizer, the other would be to simplify the problem sufficiently, so that a current-generation categorizer could learn a closed set, multiclass classification problem. With a fixed (i.e., closed) vocabulary of shapes, a fixed vocabulary of classes, and a fixed method of encoding (embedding) those shapes into a mathematical vector, a fixed set of problems, and small variations of them, could be learned.

[Kharagorgiev (2018)](#) attacked Bongard problems by simplifying them into a form that current narrow AI, particularly deep learning, could solve. He simplified the problem so that the system viewed five images in each group and had to choose between a pair of images, one of which matched the concept in the left-hand group. He further simplified the problem by pretraining a deep learning network on a million images constructed from 24 basic drawing units, including quadrangles, circles, ellipses, and polylines.  Fundamentally, he took an open-ended abstraction problem and simplified it until it was a closed classification problem.  Once simplified, the problem was easy enough for deep learning model to master.

Kharagorgiev notes "Simple off-the-shelf deep learning methods turned out to be useful for solving Bongard problems, at least in simplified classification form."  If simplified enough, then current methods are, indeed, adequate to do a passable job on these problems.  But it would then be wrong to claim then that a system that could solve the simplified problem demonstrated open-category classification.  It solved a specific set of categories with a specific set of tools.

The intelligence of Kharagorgiev's system comes largely from Kharagorgiev, himself.  He contributed the simplification, the representations, the characterization of the problem, and many other constraints to the "solution," but these are the features that would need to be contributed by the model itself if it were truly generally intelligent.  Assessing the success of a model on such a test is an evaluation of the success of the developer, not of the quality of reasoning of the model.

When a system scores well on a benchmark test by having memorized the answers from other tests, it is simplifying an ill-formed problem into a well-formed problem.  Rather than implementing general intelligence processes, it produces the correct solution to a specific problem using more conventional means.  See also Song, Yuan & Lewis, F.L. (2024), who describe a probabilistic approach to solving Bongard-type problems.

If we want a computer to be able to play chess or identify cancer in radiographs, we do not much care whether the computer solves the task in the same way as a human would (generally, they don't).  With sufficient resources, a computer could solve the mutilated checkerboard problem described earlier by evaluating every possible pattern of laying out the dominoes.  Similarly, if we narrow the scope of the problem, limiting it to only a finite class of categories, then, again, it becomes a problem that current models can solve.  If it were given the theorems of every tiling paper ever published, it could probably select the coloring argument.  In a narrow sense such a model could be said to solve the immediate problem, but it would still be wrong to say that it addresses general intelligence, and its ability to generalize beyond the immediate problem would be limited.

## Current AI models suffer from anthropogenic debt

As the proposed solution to the ARC-AGI benchmark, and as all of AI development, shows, humans play a critical role in the observed AI performance.  Humans structure the problems, choose how to represent the information, define the space of solutions, and much more.  The models themselves are much more limited in what they contribute to the solution.  Once everything is set up, parameter changes push it over the finish line.  A runner does not get credit for running a marathon if runner drives to within a few feet of the end and then walks across, yet there is a tendency to attribute intelligence to a model that solves only the final step of a problem.

Pedro Domingos (2012) noted about machine learning:

Learning = Representation + Evaluation + optimization

To make his equation even more stark, machine learning consists of three sets of numbers. One set represents the inputs to the model, one represents the outputs of the model, and one represents the parameters of the model. Machine learning consists of adjusting the parameters. Everything else is provided by humans.

In Domingos' equation, Representation and Evaluation are provided by humans. All the conceptually difficult parts of problem are provided by people, but we tend to ignore that contribution. Expert chess play was practically impossible until some human (perhaps DeGroot, 1946 or [Shannon, 1950](#)) decided to change the representation into a tree structure of moves and represent play selection as choosing branches on that tree. The models became successful when new algorithms for tree navigation and sufficient computing resources became available to traverse the tree.

From the perspective artificial general intelligence, the human contribution is anthropogenic debt. For a model to be generally intelligent or autonomous, it will have to overcome that debt and create its own representations and evaluations.

Computers are great at solving equations once they have been laid out, but so far, they are very poor at the processes needed to create novel equations. More generally, the tools that are needed to set up and represent a problem for solution are very different from those that are effective at solving an already constructed problem.

When contemporary models solve story problems, their performance can be explained by the fact that these story problems or similar ones have been included in their training set. Assessing the success of a model on such a test is an evaluation of the breadth of the training set, not an evaluation of the quality of reasoning of the model.

## Artificial general intelligence is an ill-structured problem

Part of the motivation for creating the ARC-AGI benchmark was to provide a means to assess progress toward artificial general intelligence (Chollet, 2019), to provide a feedback signal that could be used to compare systems against one another and to compare systems to humans. Later in the paper, he recognizes, however, that "optimizing for a single metric or set of metrics often leads to tradeoffs and shortcuts when it comes to everything that isn't being measured and optimized for." Those tradeoffs can be deceptive. Focusing only on the success of the model, ignores the means by which that success is achieved, and deceives us into thinking that the general problem has been solved when only the specific benchmark has been. Some people think of this ambiguity as moving the goal posts (McCorduck, 2004), but it is a tacit recognition that solving a problem does not imply that it was solved using general intelligence.

The idea of a scale against which to measure progress toward general intelligence assumes that there is some path to gradually approach it ([Morris et al., 2024](#)). Progress would consist of successively approximating the achievement endpoint of this scale. For example, if general intelligence were characterized by the number of problems a system could solve, then a scale reflecting the number of problems solved would be a good measure of progress toward general intelligence. Counting problems is not a measure of general intelligence, though, because even a pocket calculator can solve a near infinite number of problems. In any case, as an ill-structured problem, general intelligence may not be susceptible to such a scale. Progress may be discontinuous (Eldridge & Gould, 1972; Gould & Lewontin, 1979). Progress may appear to be stalled until some new invention or new discovery is found.

General intelligence requires innovations, inventions, and discoveries that have not yet been made and may never be made. The tools that we need to resolve anthropogenic debt, such as the human insight that led to the use of tree navigation for chess, will require insights that have not been imagined yet. The time to making such discoveries is also difficult to predict. There may not be some intermediate state where they partly work to yield imperfect approximations to general intelligence. They may simply be unavailable until they are.

Progress toward general intelligence will be very difficult to measure. It may not be monotonic in the sense that some innovations that appear to lead to general intelligence may eventually lead to dead ends. It may also be the case that something that looks like it is leading to general intelligence has a simpler explanation, rather than a step toward general intelligence.

It is very difficult to predict when insights will occur, and it is sometimes difficult to recognize when they have occurred. We may have ideas about the capabilities that are necessary, but it is doubtful that we will be able to check them off one by one and so measure progress. Once the necessary insights have occurred, it may then be only a small step to general intelligence.

## A theory of general intelligence is needed

There are currently many definitions of general intelligence, but no plausible theory of it. Shane Legg and Marcus Hutter (2007) list 70 definitions of intelligence, for example.

Artificial general intelligence is both an engineering challenge and a natural science challenge. The engineering challenge is to find computational methods to achieve a broad range of currently human competencies. The natural science challenge is to identify the means by which those competencies has been achieved.

Models perform at levels that may match or even exceed the competence of a certain percentage of humans (Morris et al., 2024). But we cannot infer from that demonstration of competence how it was achieved. That is a natural science questions. Experiments that are carefully constructed to test the hypothesis that a competence is general will need to be designed.

The natural science questions that we want to ask are similar to those that were investigated in the 1980s to try to distinguish between associative accounts of behavior and cognitive accounts (Lachter & Bever, 1988). The question then was whether animals or people showing a certain level of competence were doing so because they had learned specific associative rules or were using concepts (Zentall et al., 2008).

Unlike with animals or people, we know how the AI models were constructed and trained. The current GenAI models are trained to "fill in the blanks." Despite this, many claims of deep cognitive processes continue to be made even to the extent of positing that ChatGPT has a theory of mind (Kosinski, 2023). These cognitive processes are claimed to have "emerged" from the fill-in-the-blank process through scale of training volume and computational capacity. The question comes down to: Do the models actually demonstrate some cognitive capacity that would be part of general intelligence, or do they simply mimic language patterns from their training data? Given what we know about the structure, operation, and training of these models, the presence of the kind of cognitive processes that would indicate general intelligence would be an extraordinary claim and so should take extraordinary evidence. But that evidence is absent.

Also generally lacking is a theory of general intelligence.

There are several versions of an outline of a theory that conform generally to framework described by Goertzel and Pennachin (1998): "Intelligence is the maximization of a certain quantity, by a system interacting with a dynamic environment."  Their definition requires that intelligence be a well-structured problem.  Maximization of a certain quantity requires that that quantity be specified, but where does that specification come from? Why does maximization (or minimization) of that selected quantity indicate intelligence?

Space does not permit, nor do I have, a full theory of artificial general intelligence.  Instead, I will mention a few characteristics such a theory must have.  Further description is provided by Roitblat ([2020](#), [2024](#)).

**Intelligence is not behavior**.  It is not possible to build a theory consisting of just inputs (context) and outputs (behavior).  There must be latent structures and processes that mediate those relations.  The meaning of words is separate from the letter-strings that make up those words (Roitblat, 2024). GenAI models have no means to represent truth, other than that the text in their training materials is consistent, but they will need such representations.  Playing chess at a level sufficient to beat the best human players in the world is not evidence for general intelligence. There are both general and specific methods (e.g., tree traversal) that can solve it.

**How a behavior is achieved matters to generality**. Current GenAI models appear to solve many different problems with no or few examples because humans have figured out how to cast those problems into a single problem-solving framework—paraphrasing training text. If the patterns exist that can be paraphrased, then the problem can be solved. If they do not, then the model cannot solve the problem.

**There are multiple kinds of problems**.  Problems differ in scope and in the information that is needed to solve them.  They cannot all be solved with gradient descent or similar well-structured methods.  To this point most of the recent approaches to AI have focused on neural networks and on parameter adjustment.  But not all problems can be solved by parameter adjustment.  Insight problems, for example, are not solved by successively approximating a solution, but by identifying a useful way to organize and represent the problem.  To this point, there is no known method to achieve this kind of solution.  An adequate theory will need to provide a means for doing it, perhaps having to do with analogies and resonance.

**Generality refers to types of problems, not just numbers of problems**.  Pocket calculators can already solve a near infinite number of problems, but solving many problems of the same type does not make the solver general.  Rather, it must solve problems of multiple types, including well-structured, ill-structured, formal, underdetermined, fuzzy, and others.

**Anthropogenic debt must be resolved to achieve general intelligence**. Although it is rarely acknowledged, the problems that computers can solve today owe a great deal to the contribution of human designers and content contributors.  Models cannot be autonomous until they can provide the structure and representations now specified by humans.

Of these, the most immediate task that would generate substantial value would be, I think, to analyze the anthropogenic debt.  Acknowledging it and understanding just what the contribution of humans is to AI problem solving is likely to have a large impact on fear of AI, on regulations, and on investment in this area.  It would also bear on all of the other questions because much of this debt concerns types of problems that cannot be addressed through parameter adjustment.